\documentclass[10pt,amssymb,twocolumn,notitlepage]{article}

\usepackage{epigraph}
\usepackage[affil-it]{authblk}
\usepackage{dcolumn}
\usepackage[stable]{footmisc}
\usepackage{blkarray}
\usepackage{bm}
\usepackage[mathscr]{euscript}
\usepackage[font=scriptsize]{caption}
\usepackage{subcaption,graphicx}
\usepackage[table]{xcolor}
\usepackage[margin=1.7cm]{geometry}
\usepackage{tcolorbox}
\usepackage{hyperref}
\usepackage{gensymb}
\usepackage[font=footnotesize]{caption}
\usepackage{hyperref}
\hypersetup{
    colorlinks=true,
    linkcolor=blue,
    filecolor=magenta,      
    urlcolor=blue,
}
\usepackage{amssymb}
\usepackage{makecell}
\usepackage{algorithm,lipsum,xcolor,caption}
\usepackage{enumitem}
\usepackage{amsmath,esint}
\usepackage{fancyvrb}
\usepackage{xcolor}
\usepackage{arydshln}
\usepackage{mathtools}

\definecolor{newblue}{rgb}{0.0, 0.28, 0.67}
\definecolor{newgreen}{rgb}{0.13, 0.55, 0.13}
\definecolor{newred}{rgb}{0.87, 0.72, 0.53}

\newcommand{\R}{\mathbb{R}}
\newcommand{\lbar}{\{\kern-0.5ex|}
\newcommand{\rbar}{|\kern-0.5ex\}}
\usepackage{stackengine}

\definecolor{newblue}{rgb}{0.0, 0.28, 0.67}
\definecolor{newgreen}{rgb}{0.13, 0.55, 0.13}
\definecolor{newred}{rgb}{0.87, 0.72, 0.53}

\title{The Classic Cross-Correlation and the \\ 
Real-Valued Jaccard and Coincidence Indices}
\author{Luciano da Fontoura Costa \\ \emph{luciano@ifsc.usp.br}}
\affil{S\~ao Carlos Institute of Physics -- DFCM/USP} 
\date{14th Nov 2021}

\begin{document}

\twocolumn[
\begin{@twocolumnfalse}
    \maketitle
    \begin{abstract}
In this work we describe and compare the classic inner product and Pearson correlation coefficient
as well as the recently introduced real-valued Jaccard and coincidence indices.  Special attention is given to diverse schemes for taking into account the signs of the operands, as well as on the study of the geometry of the scalar field surface  related to the generalized multiset binary operations underling the considered similarity indices.  The possibility to split the classic inner product, cross-correlation, and Pearson correlation coefficient is also described.
    \end{abstract}
\end{@twocolumnfalse} 
]

\setlength{\epigraphwidth}{.49\textwidth}
\epigraph{`At the horizon line, where the sea meets the sky, countless sails.'}
{\emph{LdaFC}}

\section{Introduction}

Though not often realized, the correlation, as well as the closely related convolution
binary operators (in the mathematical sense of taking  two arguments) are among 
the most frequently employed operations in science and technology.  Basically, the
correlation between two functions $f(x)$ and $g(x)$ can be understood in terms  of 
the \emph{inner product}, which is a functional acting over the whole extent of both functions.

More specifically, the inner product can be understood as being related to the product of
one of the vectors $f$ (or functions) by the projection of the other onto $f$.  Provided
the magnitudes of the two vectors are kept constant, the inner product will also quantify
the \emph{similarity} between the two vectors, as gauged from the smallest angle between
them.

Two other similarity approaches, namely the real-valued Jaccard and coincidence indices,
have been recently proposed~\cite{CostaJaccard,CostaMset,CostaSimilarity,CostaGenMops},
mainly based on extensions of the multiset theory to take into account real, possibly negative
values.  

In the present work, we aim at studying in some detail the structure and geometry of these
three considered indices, namely the inner product as well as the real-valued Jaccard
and coincidence indices.  We start by presenting some basic concepts related to the
inner product and data standardization, which is often applied to datasets and which
implied negative respective values.   Then, we revise and present several schemes that
can be adopted to express the sign alignment between two real values (i.e.~$x y>0$ or
$x y <0$).  The several new multiset binary operations (in the sense of taking two arguments)
are then revised, which are involved in the considered similarity indices.

The real-valued Jaccard and coincidence indices are presented next, including an interesting
result relating the classic inner product with two generalized multiset operations.
The geometry and symmetry of the considered similarity indices is then
approached from their respective versions adapted to two real scalar values.  
A striking geometry is observed for the real-valued Jaccard that closely resembles the
generalized Kronecker delta function~\cite{CostaIndNeur,CostaGenMops}.

The reported concepts and methods are also employed to propose a double Pearson
correlation coefficient in which the effects of the values with same or opposite signs
can be separated and controlled as a linear combination depending on a parameter
$\alpha$, in a manner similar to that adopted for the coincidence index in~\cite{CostaCCompl}.

\section{Basic Concepts}

Given two vectors $\vec{x}$ and $\vec{y}$, both in $\R^N$, their respective inner (or scalar, or dot)
product can be expressed as:
\begin{align}
  \left< \vec{x}, \vec{y} \right> = \sum_{i=1}^N x_i \ y_i = |\vec{x}| |\vec{y}| \cos(\theta)
\end{align}

where $\theta$ is the smallest angle between the two vectors and:
\begin{align}
   |\vec{x} | =  \sqrt{ \sum_{i=1}^N x_i^2 }; \quad |\vec{y} | =  \sqrt{ \sum_{i=1}^N y_i^2 }
\end{align}

Observe that
the inner product is neither upper nor lower bound and can take positive, null
(orthogonal vectors), or negative values.

The often adopted \emph{cosine similarity} can be expressed as:
\begin{align}
  \cos(\theta) =  \frac{ \left< \vec{x}, \vec{y} \right>} { |\vec{x}| |\vec{y}|}
\end{align}

from which we see that this similarity index does not take into account the
magnitudes of the operands $x$ and $y$, but only the smallest angle
between the two vectors.  We also have that $-1 \leq \cos(\theta) \leq 1$.

Similarly, given two real functions $f(x)$ and $g(x)$, their \emph{inner product} 
is defined as:
\begin{align}
  \left< f(x), g(x) \right> = \int_{S} f(x) g(x) dx
\end{align}

where $S$ is the combined support of both functions.

When applied to real functions, the inner product allows to consider the length,
distance, orthogonality, and angle between functions in a manner analogous to that of the
inner product applied to vectors. It is also interesting to contemplate the situation
in which the inner product is applied to discretized versions of functions.    Because
functions are also vectors, in the sense of vector spaces, we will henceforth refer generically to
both vectors and functions.

It is possible to derive a distance between two operands $f$ and $g$ from the 
inner product.   First, we express the \emph{norm} of a vector $f$ in terms of the 
inner product as:
\begin{align}
  |f| = \sqrt{ \left< f(x),f (x) \right>  }
\end{align}

so that we can now define the \emph{distance} between two vectors $f$ and $g$ as:
\begin{align}
  d(f,g) = |f-g| = \sqrt{ \left< f(x) - g(x) \right> }
\end{align}

which corresponds to the \emph{Euclidean distance} between $f$ and $g$.

A \emph{complete}  inner product space is called a \emph{Hilbert space} in 
functional analysis (e.g.~\cite{KreyszigFunct}),
which can be informally understood as being an extension from vectors of linear algebra 
to real function spaces.    In addition, the concept of proximity expressed in the inner product 
also relates to topological concepts.  For instance, by \emph{complete} it is meant that all 
Cauchy sequences in the metric space has a limit contained in that same space, which 
can be informally understood as the space not having `gaps' or `holes'.

Let $X$ and $Y$ be any two random variables described by respective density
probabilities $p(x)$ and $p(y)$.

Their \emph{average} and \emph{variance} can be defined as:
\begin{align}
   \mu_ X = \int_S x p(x) dx   \\
   \mu_ Y = \int_S y p(y) dy   \\
   \sigma^2_ X = \int_S (x-\mu_X)^2 p(x) dx   \\
   \sigma^2_ Y = \int_S (y - \mu_Y)^2  p(y) dx   
\end{align}

The respective \emph{standard deviations} are:
\begin{align}
   \sigma_ X = + \sqrt{\sigma^2_X}   \\
   \sigma_ Y = + \sqrt{\sigma^2_Y}   
\end{align}

In case the joint density probability $p(x,y)$ is known,
we can define the \emph{covariance} between $X$ and $Y$ as:

Given a random variable $X$, it can be \emph{standardized} as:
\begin{equation}
  \tilde{X} = \frac{X - \mu_X}{\sigma_X}
\end{equation}

The standardization procedure is often employed, especially to make a set of random
variables more commensurate, therefore avoiding those variable with larger magnitudes
to dominate.  However, the decision to standardize or not depends on each specific
data and problem.   After standardization, each of the random variables will have zero
means and unit standard deviation.  In addition, most of the values will result inside the
interval $[-2,2]$.

The unbiased \emph{covariance} between two random variables $X$ and $Y$ represented 
in terms of respective samples $\vec{x} = x_1, x_2, \ldots, x_N$ and 
$\vec{y} = y_1, y_2, \ldots, y_N$ can be  estimated as:
\begin{equation}
  cov(X,Y) =\frac{1}{N-1} \sum_{i=1}^N \left[ x_i - \mu_X \right] \left[ y_i - \mu_Y \right] 
\end{equation}

which can be understood as a normalized inner product, in the sense that:
\begin{equation}
  cov(X,Y) =\frac{1}{N-1} \left< X - \mu_X, Y - \mu_Y \right> 
\end{equation}

Interestingly, when the cross-correlation is taken on standardized vectors $\vec{x}$
and $\vec{y}$, it becomes identical to the \emph{Pearson correlation coefficient} 
$-1 \leq P(x,y) \leq 1$.

In summary, we have seen that the classic inner product, the L2 norm, the cosine similarity, the
Euclidean distance, the cross-correlation, the covariance, and the Pearson correlation 
coefficient are all directly related to the inner product between two vector or function operands.

\section{Conjoint Sign Functions}

Given two signals $x(t)$ and $y(t)$, their respective sign functions can be expressed as:
\begin{align}
  s_x = s_x(t)  = sign(x(t))  \\
  s_y = s_y(t)  = sign(y(t))  
\end{align}

We can now define the following \emph{conjoint sign functions}:
\begin{align}
  s_p = |s_x + s_y|   \label{eq:m} \\
  s_m = |s_x - s_y|   \label{eq:s} \\
  s_{hp} = |s_x + s_y| / 2  \label{eq:m} \\
  s_{hm} = |s_x - s_y| / 2   \label{eq:s} \\
  s_{xy}  = s_x s_y   \label{eq:a}
\end{align}

\begin{figure}[h!]  
\begin{center}
   \includegraphics[width=1\linewidth]{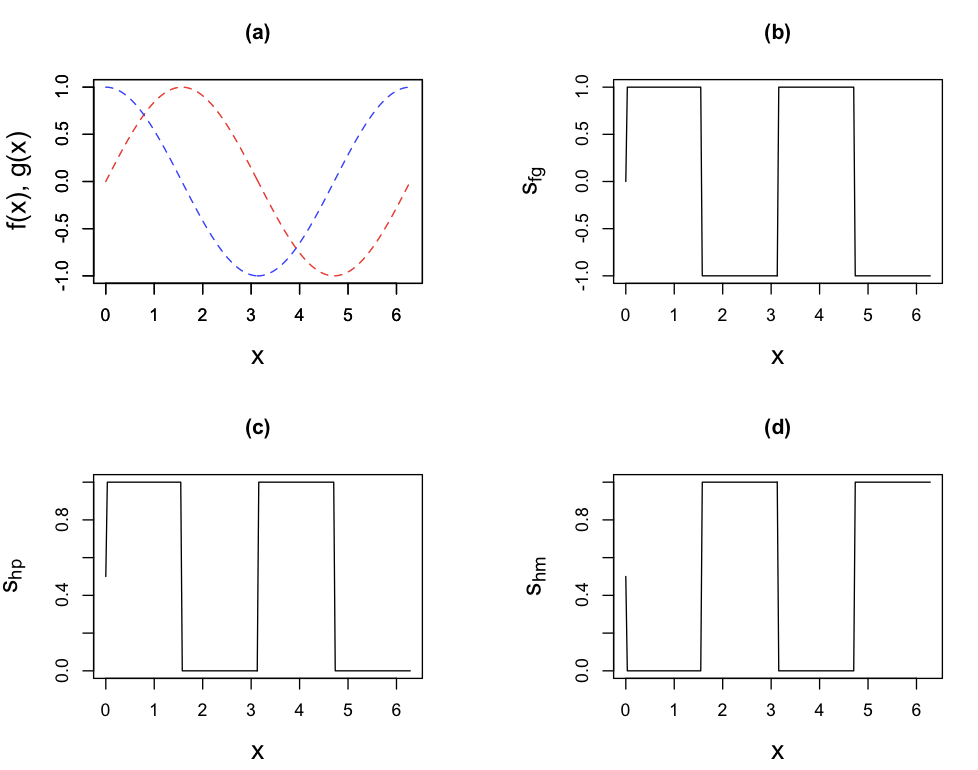}  
    \caption{Functions $f(x)$ and $g(x)$ corresponding to a whole period of sine and
    cosine, and the respective conjoint sign functions $s_{fg}$, $s_{hp}$ and $s_{hm}$. 
    Observe that $s_{fg} = s_{hp} - s_{hm}$ and also that the $y-$axis in (b-d) have different
    limits.}
    \label{fig:signs}
    \end{center}
\end{figure}

Figure~\ref{fig:signs} illustrates the functions $x_{h+}$, $x_{h-}$, and $x_{xy}$ respectively to
$x = \sin(x)$ and $y = \cos(x)$.

The function in Equation~\ref{eq:m} has been used in~\cite{mirkin},
and that in Equation~\ref{eq:a} has been employed in~\cite{Akbas1,Akbas2},
both related to the L1 norm.  The latter function has also appears in the \emph{minmod}
slope limiting function adopted in partial differential equations (e.g.~\cite{minmod}).
The function in Equation~\ref{eq:s} has been used in~\cite{CostaCCompl,CostaSimilarity}.

We also have that:
\begin{align}
  s_{hp} =  1 - s_{hm}  \\
  s_{hm} =  1 - s_{hp}  \\
  s_{xy} =  s_{p} - 1  \\
  s_{xy} =  1 - s_{m}  \\
  s_{xy} =   s_{hp} - s_{sm}  \label{eq:spm} \\
\end{align}

The generalized Kronecker delta function has been suggested~\cite{CostaSimilarity,CostaGenMops}
as a means to express not only same sign similarity as in the traditional Kronecker delta, but also
opposite sign relationships.  It can be expressed as:
\begin{align}
   \delta^{\pm}_{x,y} = \left\{ 
   \begin{array}{l}
      1 \quad \iff  x = y, \ x,y \neq 0  \\
      0 \quad \iff  x = y = 0  \\
      -1 \quad \iff  x = -y, \ x,y \neq 0  \\
    \end{array} \right.  \
\end{align}

\begin{figure}[h!]  
\begin{center}
   \includegraphics[width=0.7\linewidth]{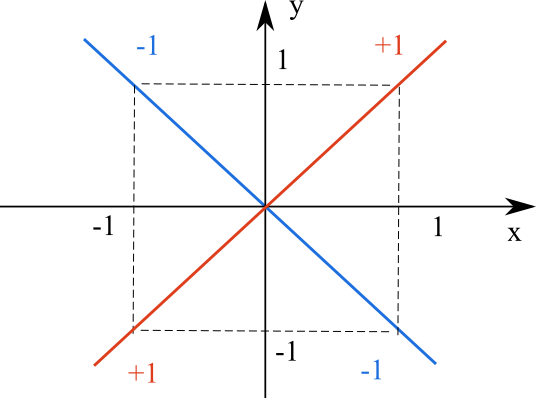}  \\ (a) \\
    \caption{The generalized Kronecker delta function allows the strictest quantification of similarities
    between values with the same or opposite signs, providing a reference for other similarity
    indices.}
    \label{fig:genKron}
    \end{center}
\end{figure}

Figure~\ref{fig:genKron} illustrates the generalized Kronecker delta function.

As developed in~\cite{CostaSimilarity}, the generalized Kronecker delta function plays a 
critically important role in defining the most strict similarity test, from which other more
tolerant similarity indices including the inner product and real-valued Jaccard index can be 
directly related.

\section{Generalized Multisets}  \label{sec:mops}

Multisets (e.g.~\cite{Hein,Knuth,Blizard,Blizard2,Thangavelu,Singh}) provide an intuitive
and interesting extension of the classic set theory so as to allow the repetition of elements.
Generalized versions of multisets~\cite{CostaJaccard,CostaMset,CostaSimilarity,CostaGenMops}
have been developed to allow real, possibly negative multiplicities.
Basically, the multiset subtraction is allowed to take negative values, so that the complement
of a multiset $A$ can be performed as $\Phi-A$, where the null multiset takes the place of the
universe set in set theory.  The generalization of multiplicities to take real, possibly
negative values~\cite{CostaJaccard,CostaMset,CostaSimilarity,CostaGenMops} allow new
binary operations to be defined between two multifunctions $x(z)$ and $y(z)$, including but not being limited
to:

\begin{align}
  f \cap g = \int_S \min\left\{ x,  y \right\} dz \label{eq:int}    \\
  f \cup g = \int_S \max\left\{ x,  y \right\} dz  \label{eq:uni}  \\
  f \sqcap g = \int_S s_{fg} \min\left\{ s_x x, s_y y \right\} dz    \label{eq:sig_int}  \\
  f \sqcup g = \int_S s_{fg} \max\left\{ s_x x, s_y y \right\} dz   \label{eq:sig_uni}   \\
  f \sqcap_- g = \int_S |s_x - s_y|/2 \min\left\{ s_x x, s_y y \right\} dz   \label{eq:half_intm}   \\
  f \sqcap_+ g = \int_S |s_x + s_y|/2 \min\left\{ s_x x, s_y y \right\} dz     \label{eq:half_intp} \\
  f \sqcup_- g = \int_S |s_x - s_y|/2 \max\left\{ s_x x, s_y y \right\} dz   \label{eq:half_unim}   \\
  f \sqcup_+ g = \int_S |s_x + s_y|/2 \max\left\{ s_x x, s_y y \right\} dz     \label{eq:half_unip} \\
  f \tilde{\sqcap} g = \int_S \min\left\{ s_x x, s_y y \right\} dz   \label{eq:abs_int}   \\
  f \tilde{\sqcup} g = \int_S  \max\left\{ s_x x, s_y y \right\} dz     \label{eq:abs_uni} 
\end{align}

Equations~\ref{eq:int} and~\ref{eq:uni} correspond to the multiset counterparts of
the set theory operations of intersection and union.  However, we do not have, as
could expected, that the intersection of a generic multiset $x$ and the null multiset $\Phi$
corresponds to the null multiset.  On the contrary, we typically have that:
\begin{equation}
  x \cap \Phi \neq \Phi
\end{equation} 

The binary operations in Equations~\ref{eq:sig_int} and~\ref{eq:sig_uni} can be understood
as the intersection and union considering negative multiplicities.  Now, we do have that:
\begin{equation}
  x \sqcap \Phi = \Phi
\end{equation} 

Equations~\ref{eq:half_intp} and~\ref{eq:half_intm} can be understood as the intersection 
operation acting only when $x$ and $y$ have the same or opposite signs. 
Equations~\ref{eq:half_unip} and~\ref{eq:half_unim} are have the analogous effect
regarding union.

Equations~\ref{eq:abs_int} and~\ref{eq:abs_uni} can be understood as the intersection and
union acting on the absolute values of multiplicities.

These several operations, which allow great flexibility for taking into account diverse
combinations of operands signs, provide the basis for obtaining the similarity indices considered in
the present work.

\section{The Real-Valued Jaccard and Coincidence Indices}

It has been described~\cite{CostaJaccard,CostaMset,CostaSimilarity,CostaGenMops} that, when 
generalized to real, possibly negative multiplicities, the Jaccard index becomes:
\begin{align}
  \mathcal{J}_R(f,g) = \frac{\int_S s_{fg} \min\left\{ sx_x, s_y y \right\} dx}   
  {\int_S  \max\left\{ sf_f, s_g g \right\} dx} = \frac{f \sqcap g}  {f \tilde{\sqcup} g}
\end{align}

which has been called the \emph{real-valued Jaccard similarity index}.

Interestingly, it can be verified that the following equation is identical to the previous one,
therefore providing an alternative definition for the real-valued Jaccard index:
\begin{align}
  \mathcal{J}_R(f,g) = \frac{\int_S \left[ f(x) \ g(x) \right]  dx}   
  { \left[ \int_S  \max\left\{ sf_f, s_g g \right\} dx \right]^2} = 
  \frac{\left< f, g \right>}  {\left[ f \tilde{\sqcup} g \right]^2}
\end{align}

Thus, we have that:
\begin{align}
  \mathcal{J}_R(f,g) =  \frac{f \sqcap g}  {f \tilde{\sqcup} g} =
  \frac{\left< f, g \right>}  {\left[ f \tilde{\sqcup} g \right]^2}
\end{align}

which then implies:
\begin{equation}
  \boxed{\left< f, g \right> =  \left[ f \sqcap g \right] \left[ f \tilde{\sqcup} g \right]}
\end{equation}

This result illustrates the flexibility of the generalized multiset operations described
in Section~\ref{sec:mops}.  In addition, it establishes an important link between the
real-valued Jaccard, as well as the coincidence indices, with the classic inner product.
As a matter of fact, this result shows that the cross correlation, which consists in the
successive sliding application of the inner product, can actually be used for 
calculation of the real-valued Jaccard index, and vice versa, provided the proper
normalization is taken into account.  This important link will be further considered
in Section~\ref{sec:geometry} in order to better understand the properties of the
similarity indices considered in this work.

Given that the Jaccard index is not capable of taking into account the relative interiority
between the two compared sets, vectors or functions~\cite{CostaJaccard}, it has been complemented by
considering the interiority index (also overlap~\cite{Kavitha}) which, when adapted to
real, possibly negative values yields:
\begin{align}
  \mathcal{I}_R(f,g) = \frac{\int_S \min\left\{ sx_x, s_y y \right\} dx}   
  {\min\left\{ S_f, S_g \right\}} = \frac{f \sqcap g}  { \min\left\{ S_f, S_g \right\}}
\end{align}

where:
\begin{align}
    S_f = \int_S s_f f(x) dx \\
    S_g = \int_S s_g g(x) dx 
\end{align}

The \emph{coincidence index} can now be expressed~\cite{CostaJaccard,CostaMset,CostaSimilarity,CostaGenMops} as corresponding to
the product between the real-valued Jaccard and interiority indices:
\begin{align}
  \mathcal{C}_R(f,g) = 
    \frac{ \left[ f \sqcap g \right] \left[ f \tilde{\sqcap} g \right]}   { \left[ f \tilde{\sqcup} g \right] \min\left\{ S_f, S_g \right\}}
\end{align}

or, in expanded notation:
\begin{align}
  &\mathcal{C}_R(f,g) =  \nonumber \\
  &\quad = \frac{  \left[ \int_S s_{fg} \min\left\{ sx_x, s_y y \right\} dx \right] \left[ \int_S \min\left\{ sx_x, s_y y \right\} dx \right] }  
   {\left[ \int_S  \max\left\{ sf_f, s_g g \right\} dx \right] \left[ \min\left\{ S_f, S_g \right\} \right]} 
\end{align}

\section{The Geometry of Similarity} \label{sec:geometry}

In this section we study in some detail the geometry and symmetries of the above
presented similarity indices in order to better understand the properties and effects
of each index.

We start by considering the real-valued Jaccard similarity index 
rewritten for two real scalar values $f=x$ and $g=y$:
\begin{align}
  &\mathcal{J}_R(x,y) = \frac{s_{xy} \min\left\{  |x|, |y| \right\}}  
     {  \max\left\{ s_x x , s_y y \right\}  }
\end{align}

We can then separate the numerator and denominator as:
\begin{align}
   &A_1 (x,y) =  s_{xy}  \min\left\{  s_x x, s_y y \right\} = x \sqcap y \label{eq:A1} \\
   &A_2 (x,y) =   \max\left\{ s_x x , s_y y \right\} = x \tilde{\sqcup} y \label{eq:A2} 
\end{align}

which, as two scalar fields on the $(x,y)$ domain, can be visualized.

We will also consider the following additional fields, directly related to
the inner product:
\begin{align}
   &A_3 (x,y) =    x \ y  \label{eq:A3} \\
   &A_4 (x,y) =    \left[ \max\left\{ s_x x , s_y y \right\} \right]^2 = \left[x \tilde{\sqcup} y \right]^2\label{eq:A4} 
\end{align}

as well as the multiset operation:
\begin{align}
   &A_5 (x,y) =    \min\left\{ s_x x , s_y y \right\}  = x \tilde{\sqcap} y  \label{eq:A5} 
\end{align}

When rewritten for scalar real values $x$ and $y$,  we have that 
$f \tilde{\sqcap} g = \min\left\{ S_x, S_y \right\} $ and therefore the coincidence index becomes:
\begin{align}
  \mathcal{C}_R(x,y) = 
    \frac{ \left[ x \sqcap y \right] \left[ x \tilde{\sqcap} y \right]}  
     { \left[ x \tilde{\sqcup} y \right]\left[ x \tilde{\sqcap} y \right]} = 
    \frac{  x \sqcap y }  { x \tilde{\sqcup} y }  = \mathcal{J}_R(x,y)
\end{align}

However, observe that this will not, in general, be the case with higher dimensional
vector operands, in which case the real-valued Jaccard and coincidence index will
be typically distinct.

\begin{figure}[h!]  
\begin{center}
   \includegraphics[width=0.7\linewidth]{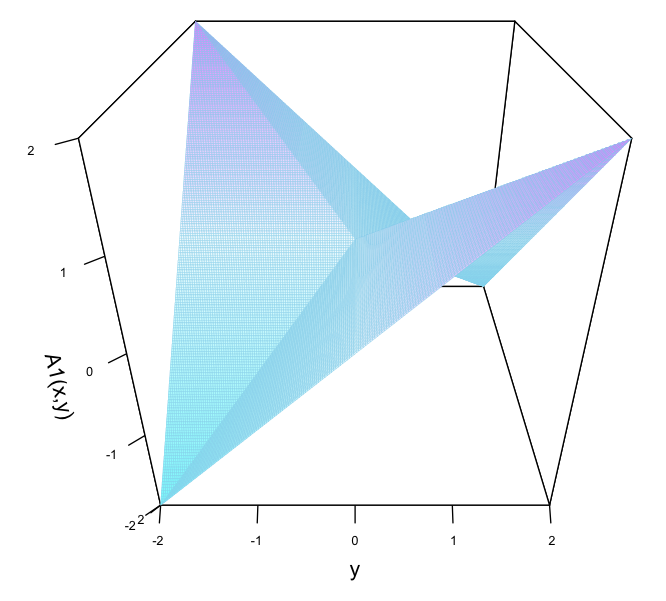}  \\ (a) \\
   \includegraphics[width=0.7\linewidth]{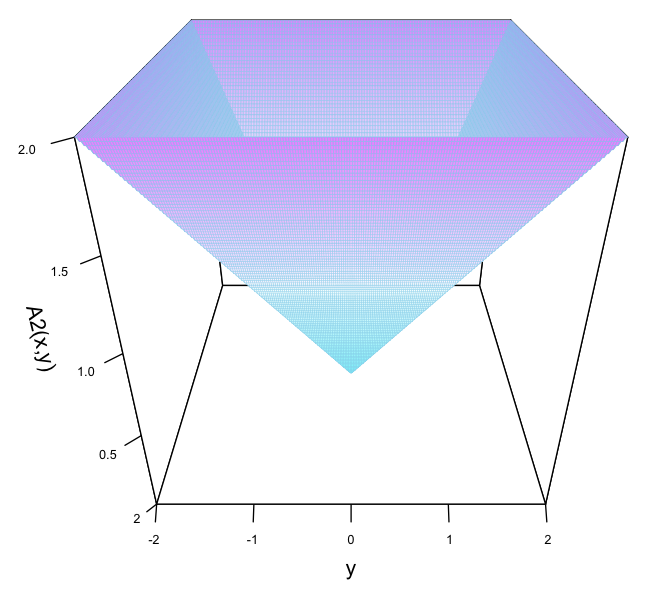}  \\ (b) \\
   \includegraphics[width=0.7\linewidth]{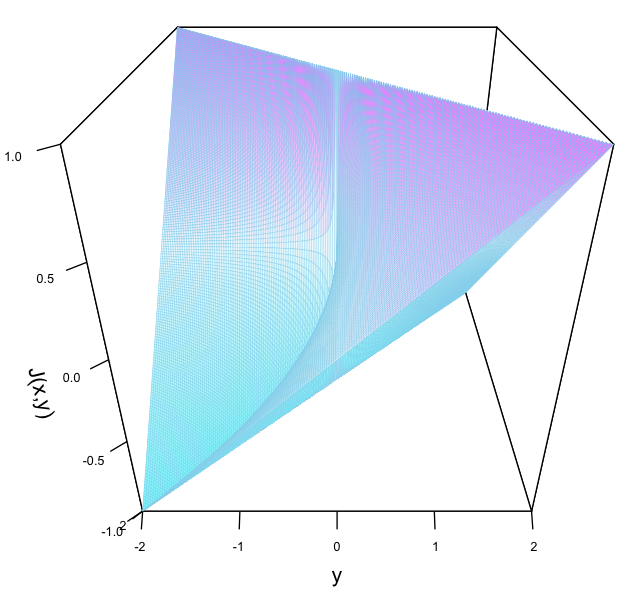}   \\ (c) \\
    \caption{The scalar fields $A_1(x,y)$ in Equation~\ref{eq:A1} (a);
    $A_2(x,y)$ in Equation~\ref{eq:A2} (b); and the real-valued
    Jaccard index $\mathcal{J}_R(x,y) = A_1(x,y)/A_2(x,y$
    (c) for $-2 \leq x \leq 2$ and $-2 \leq y \leq 2$.
    Observe the striking geometry of the real-valued Jaccard index
    shown in (c), which is closely related to the 
    generalized Kronecker delta function.}
    \label{fig:A1A2}
    \end{center}
\end{figure}

Figure~\ref{fig:A1A2} illustrates the fields $A_1(x,y)$ (a)  and $A_2(x,y)$  (b) for
$-2 \leq x \leq 2$ and $-2 \leq y \leq 2$.  

Observe the striking geometry of the surface in Figure~\ref{fig:A1A2}(c).
As a more detailed verification will reveal, this function resembles closely
the generalized Kronecker delta function~\cite{CostaIndNeur}.  Observe the
gradual, linear rotation from the identity line $x=y$ to the anti-identity
line $x=-y$.  It is this specific geometry of the real-valued Jaccard index
that contributes to enhanced performance for template matching observed
in~\cite{CostaComparing,CostaIndNeur}.

Figure~\ref{fig:A3A4} presents the multiset operations in Equations~\ref{eq:A3} and~\ref{eq:A4},
the former of which being directly related to the inner-product.

\begin{figure}[h!]  
\begin{center}
   \includegraphics[width=0.7\linewidth]{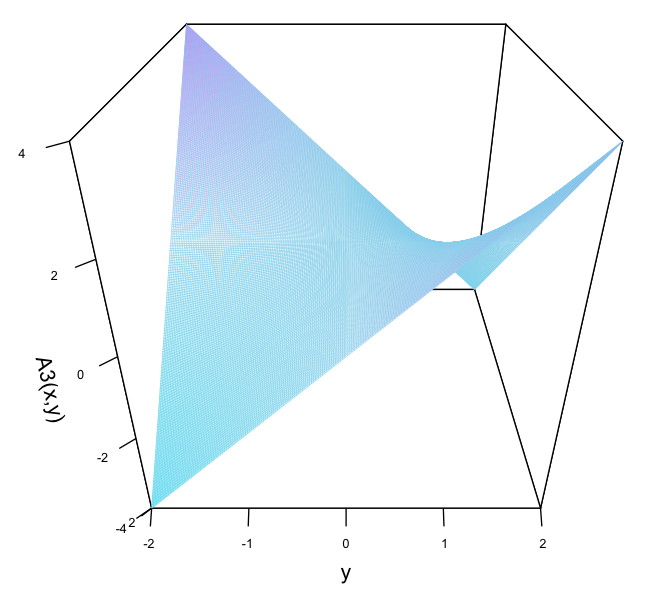}  \\ (a) \\
   \includegraphics[width=0.7\linewidth]{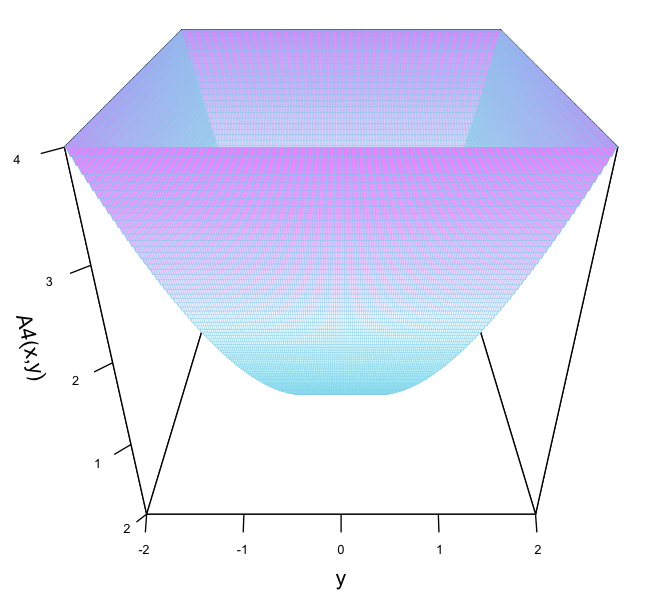}  \\ (b) 
    \caption{The scalar fields $A_3(x,y)$ in Equation~\ref{eq:A4} (a);
    $A_2(x,y)$ in Equation~\ref{eq:A2} (b);
    for $-2 \leq x \leq 2$ and $-2 \leq y \leq 2$. Recall 
    that, for scalar values, $\mathcal{C}_R(x,y) = 
    \mathcal{J}_R(x,y) = A_3(x,y)/A_4(x,y) = A_1(x,y)/A_2(x,y)$,
    which is shown in Fig.~\ref{fig:A1A2}(c).}
    \label{fig:A3A4}
    \end{center}
\end{figure}

\begin{figure}[h!]  
\begin{center}
   \includegraphics[width=0.7\linewidth]{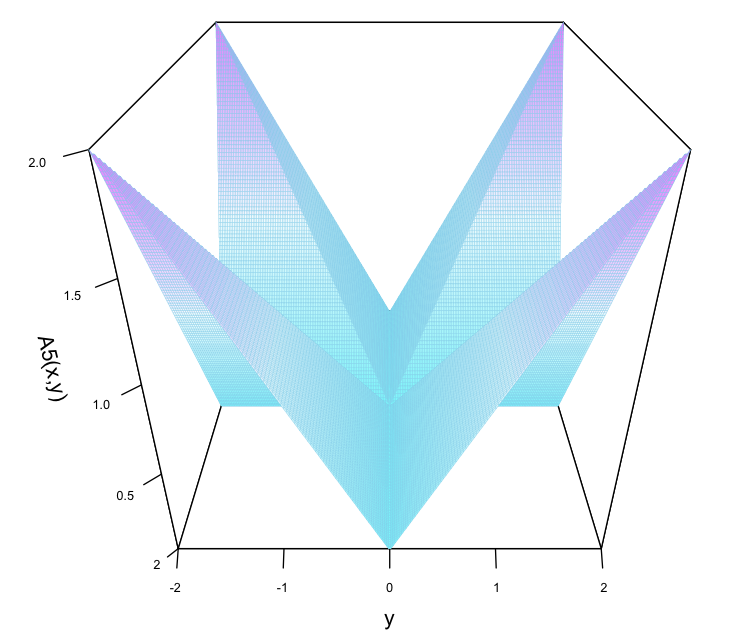}  
    \caption{The scalar field obtained for the multiset operation  $x \tilde{\sqcap} y$,
    where $x$ and $y$ are real scalar values,  for $-2 \leq x \leq 2$ and $-2 \leq y \leq 2$. }
    \label{fig:A5}
    \end{center}
\end{figure}

Figure~\ref{fig:A5} depicts the multiset absolute union operation $x \tilde{\sqcap} y$,
which plays an important role in both the interiority and coincidence indices.

We are now in position to analyze more closely the geometry of the cross-correlation and
real-valued Jaccard index.  Given the quadrant symmetries of these indices, we henceforth
focus our attention only on the first quadrant, $x\geq0$ and $y \geq 0$.

Let's consider the real-valued Jaccard similarity applied to real scalar values.  We have 
the following situations:
\begin{equation}
   \left\{ 
   \begin{array}{l}
      x > y \quad \iff \mathcal{J}_R(x,y) = \frac{y}{x}  \\
      x = y = 0 \quad \iff \mathcal{J}_R(x,y) = 0 \\  
      x < y \quad \iff \mathcal{J}_R(x,y) = \frac{x}{y}  \\
    \end{array} \right.  
\end{equation}

Given the involved symmetries, we can further restrict our attention to $x>y$.  

\begin{figure}[h!]  
\begin{center}
   \includegraphics[width=0.5\linewidth]{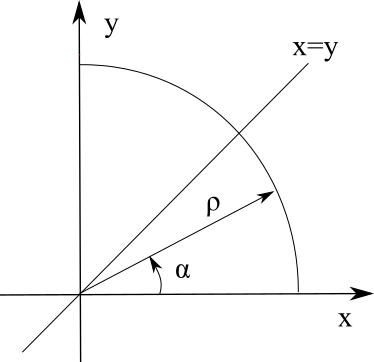}  
    \caption{The geometrical construction adopted for studying the geometry of the
    scalar field defined by the real-valued Jaccard similarity index assuming scalar
    operands in the case of $x>y$.}
    \label{fig:A3A4}
    \end{center}
\end{figure}

If we fix our attention to the points $(x,y)$, with $x>y$, so that $\rho = \sqrt{x^2 +y^2}$
is equal to a fixed constant $\tilde{\rho}$, we will find that:
\begin{align}
   \mathcal{J}_R(\alpha, \rho=\tilde{\rho}) =  \frac{y}{x} = \tan \left(\alpha \right) 
\end{align}

Therefore, we have that the real-valued Jaccard similarity along the semi-circle
defined by $\alpha$ and $\tilde{\rho}$ increases with $\tan{\alpha}$ as we go from
$\alpha = 0$ to $\alpha = \pi/4$.  At $\alpha = \pi/4$, we have $\mathcal{J}_R(x,y) = \tan(\pi/4) = 1$,
which corresponds to the classic Kronecker delta function.

The above geometry analysis reveals that the real-valued Jaccard similarity index corresponds
to a version of the Kronecker delta function in which the similarity decreases with the tangent
of $\alpha$ as one rotates from the maximum crest corresponding to the classic Kronecker
delta function.  As such, the real-valued Jaccard indeed implements a more strict quantification
of the similarities, as described in~\cite{CostaJaccard,CostaSimilarity,CostaGenMops}.

It is also interesting to consider the possibility of having:
\begin{align}
  \mathcal{J}_R(f,g) =  \frac{\left[ f \sqcap g \right]^D}  {f \tilde{\sqcup} g} 
\end{align}

\begin{figure}[h!]  
\begin{center}
   \includegraphics[width=0.55\linewidth]{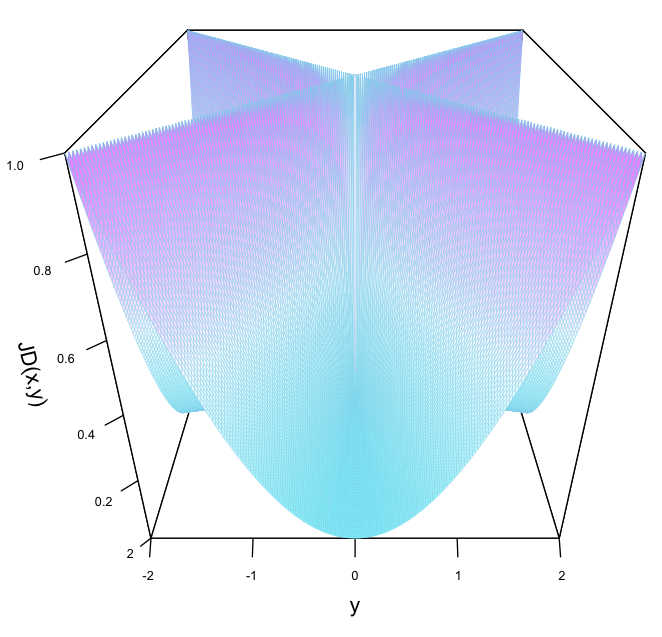}  \\ (a) \\
   \includegraphics[width=0.55\linewidth]{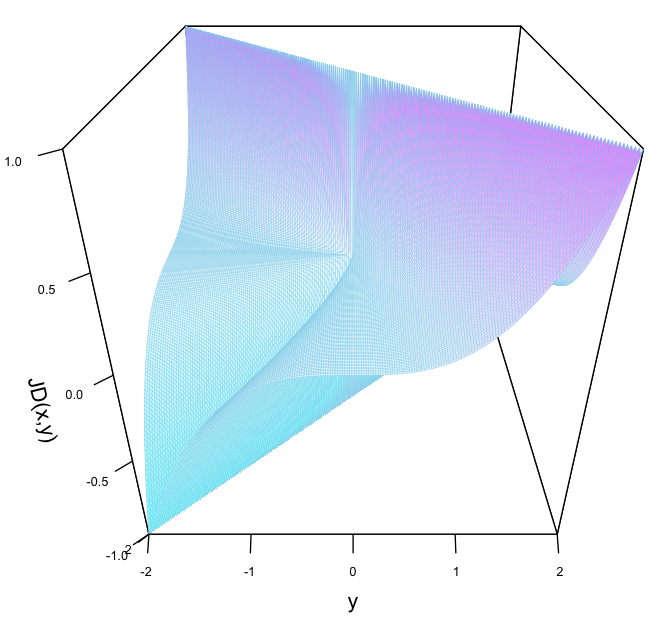}  \\ (b) \\
   \includegraphics[width=0.55\linewidth]{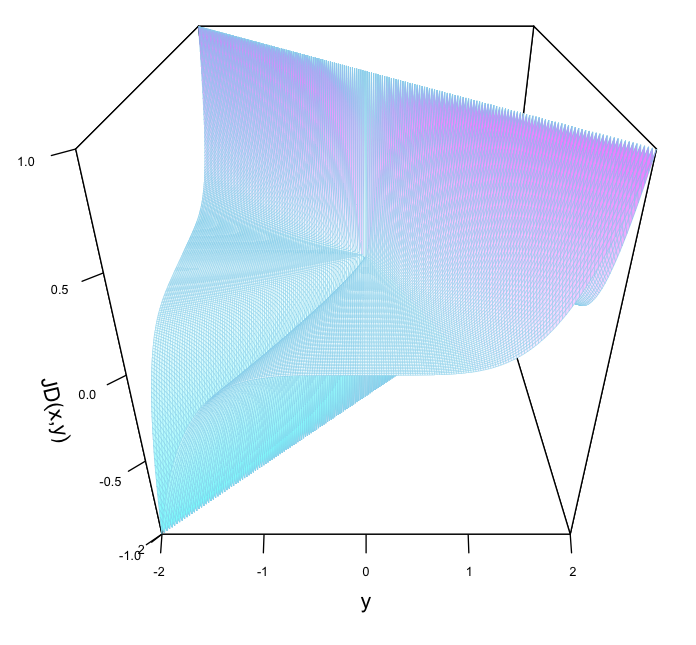}  \\ (c) \\
   \includegraphics[width=0.55\linewidth]{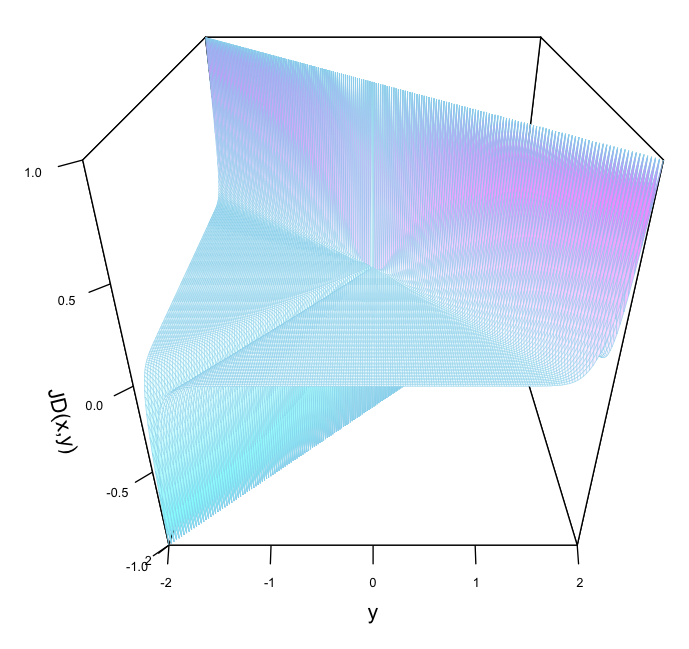}  \\ (d) \\
    \caption{The scala version of the real-valued Jaccard index with the numerator
    taken to the power of $D$ for $D = 2$ (a), $3$, (b), $5$ (c), and $21$ (d).
    Observe the convergence of this index to the generalized Kronecker delta 
    function as $ D \rightarrow \infty$, $D$ odd.}
    \label{fig:Ds}
    \end{center}
\end{figure}

Figure~\ref{fig:Ds} illustrates the scalar versions of the real-valued Jaccard index
with the numerator taken to the power of $D$ for $D = 2, 3, 5$ and $21$.  Two
particularly interesting properties can be observed.  First, we have that even
values of $D$ will imply the real-valued Jaccard index to become related to
the  absolute value of the generalized Kronecker delta function, which can be
of interest for certain applications.  Second, it is interesting to observe that the
real-valued Jaccard index converges to the generalized Kronecker delta function
as $D \rightarrow \infty$, for $D$ odd.

The above results indicate that the poser $D$ controls how much the real-valued
Jaccard index is strict regarding the quantification of similarity.   More strict similarity
quantifications will be characterized by steeper crests in the respectively obtained
geometries.  A similar result is obsered for the case of the coincidence index.

\section{The Double Pearson Coefficient}

As described recently in~\cite{CostaCCompl}, it is interesting to split Equation~\ref{eq:sig_int}
as a combination of the indices proposed in~\cite{mirkin}, i.e.:
\begin{equation}
    f \sqcap g= 2 \left[\alpha \right] \left[ f \sqcap_+ g \right] - 2 \left[1- \alpha \right] \left[ f \sqcap_- g \right]
\end{equation}

where $0 \leq \alpha \leq 1$ controls the contribution of the pairs of values $x$ and $y$ that
have the same or opposite signs on the resulting integration.   This resource has proven to
allow an effective means for obtaining progressions of datasets represented as complex
networks that are increasingly more connected for increasing values of $\alpha$.

The concepts and methods reviewed and reported in this work paves the way to obtaining 
an analogous decomposition of the classic inner product as well as its normalized version
known as the Pearson correlation coefficient.

Let's define the functionals:
\begin{align}
  \left< f, g \right>_- = \int_S |s_x - s_y|/2 \ x y \ dz   \label{eq:innerm}   \\
  \left< f, g \right>_+ = \int_S |s_x + s_y|/2 \ x  y \ dz   \label{eq:innerp}   \\
\end{align}

The double inner product can therefore be written as:
\begin{equation}
    \left< f, g \right> = \left[\alpha \right] \left<f, g\right>_- - \left[1- \alpha \right]  \left<f, g\right>_+ 
\end{equation}

It may also be interesting to consider the two split terms separately in order to provide
additional information about the joint variation of the two operands $x$ and $y$.

\begin{figure}[h!]  
\begin{center}
   \includegraphics[width=0.9\linewidth]{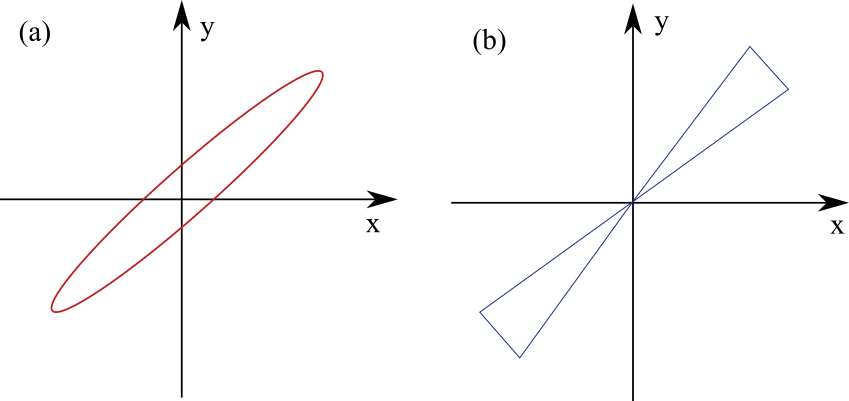}  
    \caption{Two rather distinct point distributions which could have the same inner product
    or Pearson correlation coefficient.  The consideration of the double Pearson correlation
    suggested in this work allows these situations to be effectively distinguished.  For instance,
    in the case of this specific example, we would have that $\left<f, g\right>_-$ would be non-zero
    at $(x=0,y=0)$ for (a) and zero for the point distribution in (b). }
    \label{fig:A3A4}
    \end{center}
\end{figure}

\section{Concluding Remarks}

The prsent work has addressed the properties of the inner product as well as the 
real-valued Jaccard and coincidence indices, with emphasis on the several
interesting schemes that can be employed to take into account the sign of the
operands.  These developments are strongly based on the generalization of
multisets to take into account real, possibly negative values.  In particular,
these generalizations allow several additional binary operators between 
multisets to be defined, many of which have been presented here.
 
Special attention has been give to the geometrical characterization
of the surfaces arising by the several involved multiset operators which,
for scalar values of the operands $x$ and $y$, define respective scalar
fields that can be conveniently visualized.  This approach has allowed us
to observe that the real-valued Jaccard and coincidence indices present
a geometry that closely resembles the generalized Kronecker delta function,
involving a rotation from the identity line $x=y$ to the anti-identity
line $x=-y$ following the tangent function.  It has also been verified that taking
the numerator of the real-valued Jaccard index to the power of $D$, with $D$ odd,
provides an effective manner to control de degree of how much strict the similarity
quantification is performed.  In particular, these functions converge to the
generalized Kronecker delta as $D \rightarrow \infty$, $D$ odd.

The reported approach also motivates the consideration of local and global
properties of the geometry of the obtained surfaces, such as the respective
gradients, as a means to formally specify criteria for the similarity quantification.
For instance, one may aim at achieving minimum variation of the gradient
magnitude as one moves from  $x=y$ to $x=-y$.  

The described concepts also paved the way to developing a double Pearson
correlation coefficient, in which the contribution of the values $x$ and $y$ with
the same or opposite signs can be separated and taken as a linear combination
controlled by a respective parameter $0 \leq \alpha \leq 1$, in a similar manner
to that described recently in~\cite{CostaCCompl,CostaCaleidoscope}.

\vspace{0.7cm}
\emph{Acknowledgments.}

Luciano da F. Costa
thanks CNPq (grant no.~307085/2018-0) and FAPESP (grant 15/22308-2).  
\vspace{1cm}

\bibliography{mybib}
\bibliographystyle{unsrt}

\end{document}